\documentclass[graybox]{svmult}

\usepackage{mathptmx}       
\usepackage{helvet}         
\usepackage{courier}        
\usepackage{makeidx}         
\usepackage{graphicx}        
\usepackage{multicol}        
\usepackage[bottom]{footmisc}

\usepackage{url}

\usepackage{subfigure}
\makeindex             

\begin{document}

\graphicspath{{figs/}}

\title*{Maze solvers demystified and some other thoughts \thanks{This is a preliminary version of the chapter to be published in Adamatzky A. (Ed.) Shortest path solvers. From software to wetware. Springer, 2018. }}
\author{Andrew Adamatzky}
\institute{Andrew Adamatzky \at Unconventional Computing Lab, UWE, Bristol, BS16 1QY, UK, \email{andrew.adamatzky@uwe.ac.uk}}

\maketitle

\abstract*{There is a growing interest towards implementation of maze solving in spatially-extended physical, chemical and living systems. Several reports of prototypes  attracted great publicity, e.g. maze solving with slime mould and epithelial cells, maze navigating droplets.  We show that most prototypes utilise one of two phenomena: a shortest  path in a maze is a path of the least resistance for fluid and current flow, and  a shortest path is a path of the steepest gradient of chemoattractants.  We discuss that substrates with so-called maze-solving capabilities simply trace flow currents or chemical diffusion gradients. We illustrate our thoughts with a model of flow and experiments with slime mould. The chapter ends with a discussion of experiments on maze solving with plant roots and leeches which show limitations of the chemical diffusion maze-solving approach.  
}

\abstract{There is a growing interest towards implementation of maze solving in spatially-extended physical, chemical and living systems. Several reports of prototypes  attracted great publicity, e.g. maze solving with slime mould and epithelial cells, maze navigating droplets.  We show that most prototypes utilise one of two phenomena: a shortest  path in a maze is a path of the least resistance for fluid and current flow, and  a shortest path is a path of the steepest gradient of chemoattractants.  We discuss that substrates with so-caflled maze-solving capabilities simply trace flow currents or chemical diffusion gradients. We illustrate our thoughts with a model of flow and experiments with slime mould. The chapter ends with a discussion of experiments on maze solving with plant roots and leeches which show limitations of the chemical diffusion maze-solving approach }


\section{Introduction}

\begin{figure}[!tbp]
\centering
\subfigure[]{\includegraphics[width=0.99\textwidth]{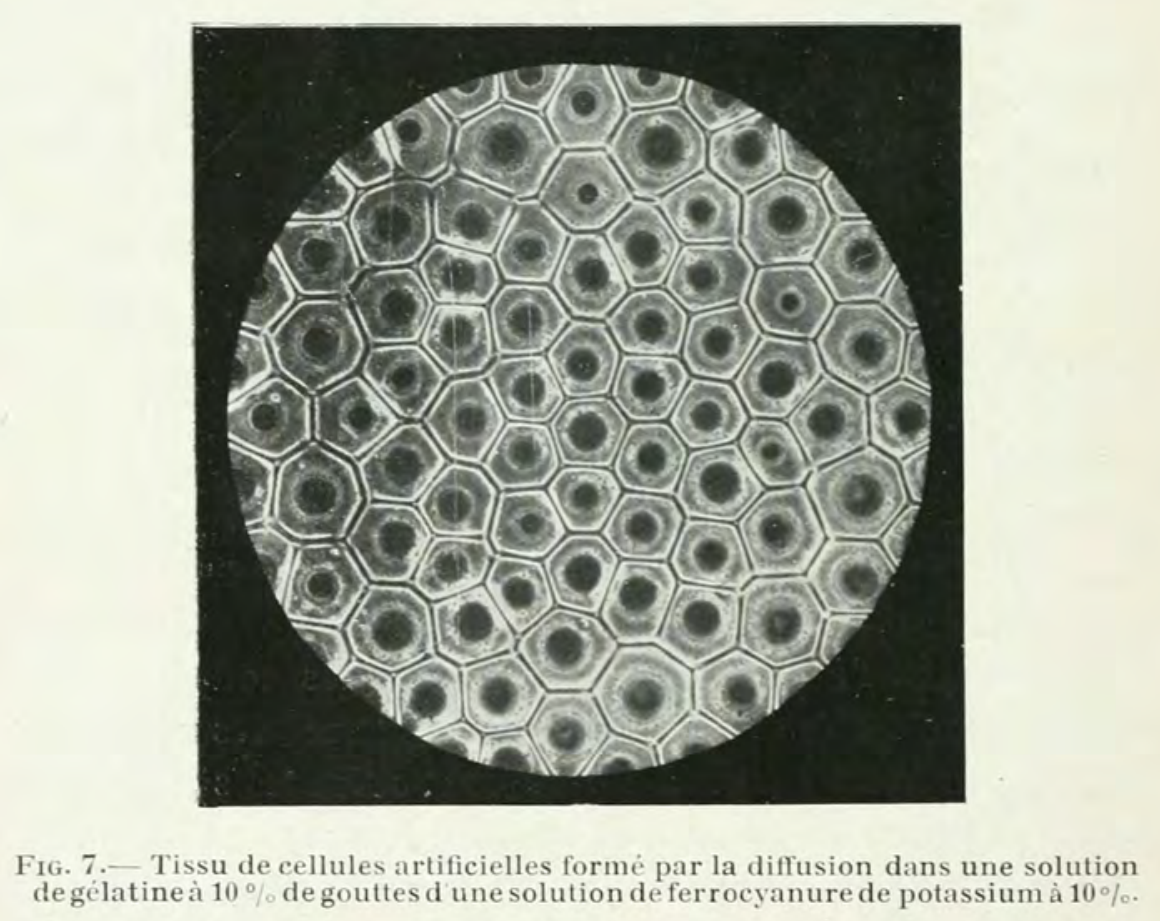}}
\subfigure[]{\includegraphics[width=0.35\textwidth]{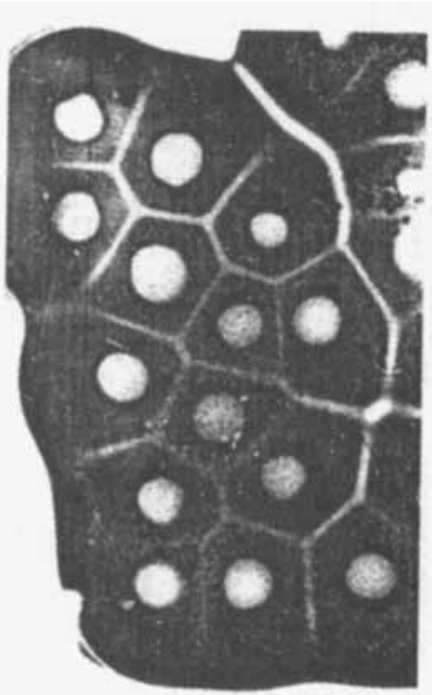}}
\subfigure[]{\includegraphics[width=0.35\textwidth]{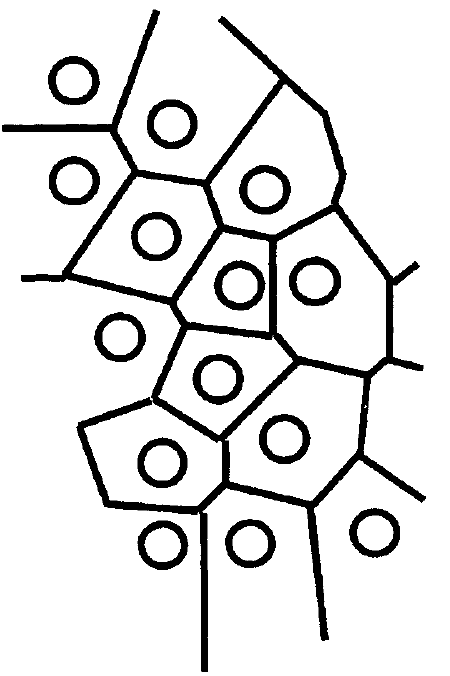}}
\caption{Unconventional computing is an art of interpretation. 
(a)~Cellular structure produced by St{\'e}phane Leduc in 1910 with drops of potassium ferrocyanide diffusing in gelatine ~\cite{leduc1910theorie}.
(bc)~Chemical processor made by Adamatzky and Tolmachiev in 1996~\cite{tolmachiev1996chemical}: photo of a completed reaction (b) and corresponding Voronoi diagram (c).}
\label{leducVD}
\end{figure}

To solve a maze\footnote{A labyrinth is a maze with  a single path to an exit/destination} is to find a route from the source site to the destination site.  In \cite{adamatzky2017physical} we reviewed  experimental laboratory prototypes of maze solvers. We speculated that the experimental laboratory prototypes of maze solvers, despite looking different, use the same principles in their actions: mapping and tracing. A maze is mapped in parallel by developing chemical, electrical, or thermal gradients\footnote{This is a material implementation of 1961 Lee algorithm, where  each site of a maze gets a label showing a number of steps someone must make to reach the site from the destination site~\cite{lee1961algorithm, rubin1974lee}}.  A path from a given source site to the destination site is traced in the mapped maze using living cells, fluid flows or electrical current. The traced paths are visualised with morphological structures of living cells, dyes, droplets, thermal sensing or glow-charge. The experimental laboratory maze solvers vary in their speeds substantially. The solvers based on glow-discharge~\cite{reyes2002glow, dubinov2014glow} or thermal visualisation of  a path~\cite{ayrinhac2014electric}, and the solver utilising crystallisation~\cite{adamatzky2009hot} produce the traced path in a matter of milliseconds or seconds. Prototypes employing assembly of conductive particles~\cite{nair2015maze}, dyes~\cite{fuerstman2003solving}, droplets~\cite{lagzi2010maze,cejkova2014dynamics}
and waves~\cite{agladze1997finding,adamatzky2002collision} give us results in minutes. Living creatures --- slime mould~\cite{adamatzky2012slime} and epithelial cells~\cite{scherber2012epithelial} --- require hours or days to trace the path.  

Chemical, physical and living maze solvers are conventional examples of unconventional computers. In the present chapter we do not provide all technical details of the experimental laboratory prototypes, these can be found in  \cite{adamatzky2017physical}, but rather share our thoughts  on maze solvers in a context of  unconventional computing and discuss some experiments with inconclusive results. 

 Whilst mentioning `unconventional computing' we might provide a definition of the field. The field is vaguely defined as the computing with physical, chemical and living substrates (as if conventional computers compute with `non-physical' substrates!). In our recent opinion paper~\cite{adamatzky2017east} unconventional computists provided several definitions, e.g.  challenging impossibilities (Cristian Calude), going beyond discriminative knowledge (Kenichi Morita), intrinsic parallelism and nonuniversality (Selim Akl), and continuous computation (Bruce MacLennan).  Jos\'{e} F\'{e}lix Costa defines the unconventional computing as `physics of measurement' which echoes with our own opinion of the unconventional computing as an art of interpretation~\cite{adamatzky2010physarum}. Take, for example, the famous, and still very much relevant, book by  
St{\'e}phane Leduc  ``Th{\'e}orie physico-chimique de la vie et g{\'e}n{\'e}rations spontan{\'e}es'' published in Paris in 1910. Not only did this book lay a foundation of the Artificial Life but somewhat contributed to the field of unconventional computing. Namely, have a look at the Fig.~\ref{leducVD}a. This is a structure that emerged when Leduc placed drops on potassium ferrocyanide on the gelatine gel. Neighbouring diffusing drops applied pressure to each other and diffusion stopped at the bisectors between the drops. Leduc presented this as a chemical model of multi-cellular formation. Unaware of the Leduc's experiments Adamatzky and Tolmachiev rediscovered a similar formation in 1996 and reinterpreted it as a chemical processor which computes Voronoi diagram of a planar set of points~\cite{tolmachiev1996chemical}: the data points are represented by drops of potassium iodide diffusing in a thin-layer agar with palladium chloride (Fig.~\ref{leducVD}bc). These our historical reminiscences smoothly flow into the next section of the chapter on fluid mappers.

\section{Fluid mappers. Shortest path is a path of the least hydrodynamic resistance}

\begin{figure}[!tbp]
\centering
\includegraphics[width=0.95\textwidth]{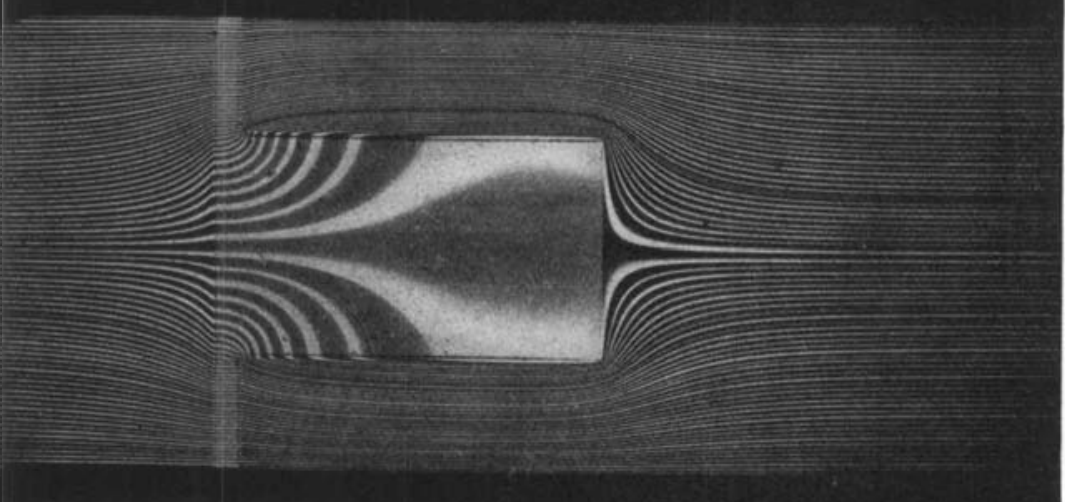}
\caption{Stream-lines of a fluid flow around a domain with low permeability, designs were proposed in 1904. From ~\cite{hele1905hydrodynamical}.}
\label{heleshaw}
\end{figure}

In 1900 Hele-Shaw and Hay developed an analogy between stream-lines of a fluid flow in a thin layer and the lines of magnetic induction in a uniform magnetic field~\cite{hele1900lines}: 
pressure gradient of a fluid flow is equivalent to magnetic intensity and rate of the flow is analogous to magnetic induction. As Hele-Shaw and Hay wrote~\cite{hele1900lines}:
\begin{quote}
The method described is the only one hitherto known which enables us to determined the lines of induction in the substance of a solid magnetic body. 
\end{quote}
In 1904 they applied their approach to solve a ``problem of the magnetic flux distortion brought about by armature teeth''~\cite{hele1905hydrodynamical} (Fig.~\ref{heleshaw}).

\begin{figure}[!tbp]
\centering
\includegraphics[width=0.64\textwidth]{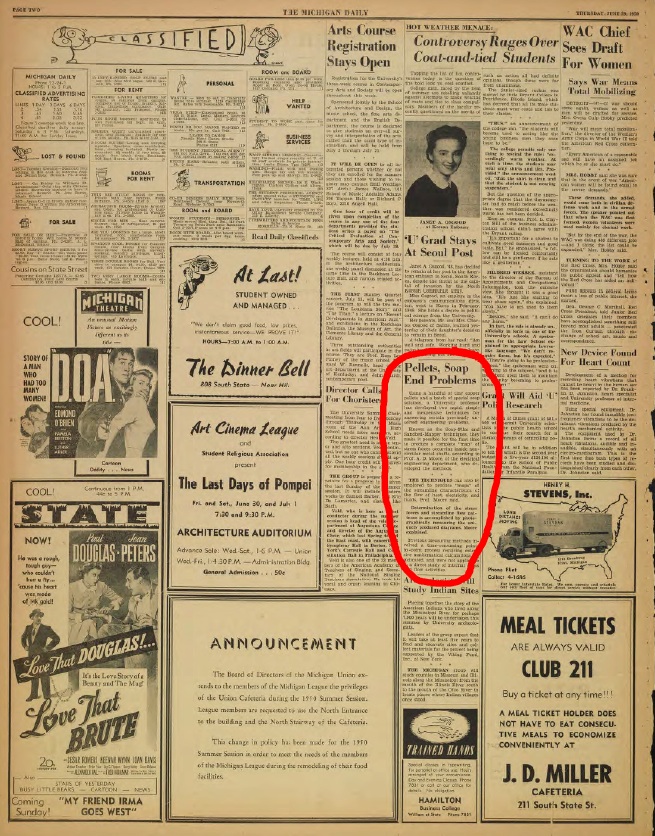}
\includegraphics[width=0.31\textwidth]{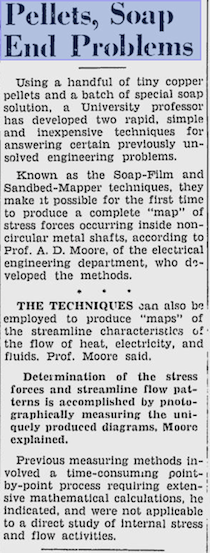}
\caption{A short news story about Moore's works was published in `` The Michigan Daily'' newspaper on June 29, 1950~\cite{michigandaily29061950}. \copyright The Michigan Daily }
\label{michigandaily}
\end{figure}

Hele-Shaw and Hay 's idea was picked up by Arthur Dearth Moore who developed fluid flow mapping devices~\cite{moore1949fields} (Fig.~\ref{michigandaily}).\footnote{Moore has also invented hydrocal, a hydraulic computing device for solving unsteady  problem in heat transfer~\cite{moore1936hydrocal} at the same time when Luk'yanov's invented his famous hydraulic differential equations solver~\cite{luk1939hydraulic}}.  The Moore's fluid mapper is made of a cast slab, covered by a glass plate, with input (source) and output (sink) ports, fluid flow lines are visualised by traces from dissolving crystals of  potassium permanganate  or methylene blue. He shown that his fluid mappers can simulate electrostatic and magnetic fields, electric current, heat transfer and chemical diffusion~\cite{moore1949fields}. This is a description of the mapper in Moore's own words~\cite{moore1955fluid}:
\begin{quote}
When a given potential field situation is to be portrayed, the lower member
of the fluid mapper is built to scale, with suitable boundaries, open or closed;
islands, if any; one or more sources or sinks; and so on. Each source or sink
is connected by a rubber tube to a tank, so that raising or lowering a tank will
induce flow in the flow space. When the operation is conducted so that the
flow is not affected by inertia, the flow pattern set up can quite accurately
duplicate either the equipotential lines, or else the flux lines, of the potential
field under consideration.
\end{quote}
Moore mentioned `islands', which could play a role of obstacles or even maze walls, when a collision-free shortest path is calculated or a maze solved, however, there is no published evidence that Moore applied his inventions to solve mazes. Maybe he did. The fluid mappers became popular, for a decade, and have been used to solve engineering problems of underground gas recovery and canal 
seepage~\footnote{\url{http://quod.lib.umich.edu/b/bhlead/umich-bhl-851959?rgn=main;view=text}}.

\begin{figure}[!tbp]
\centering
\subfigure[]{\includegraphics[width=0.5\textwidth]{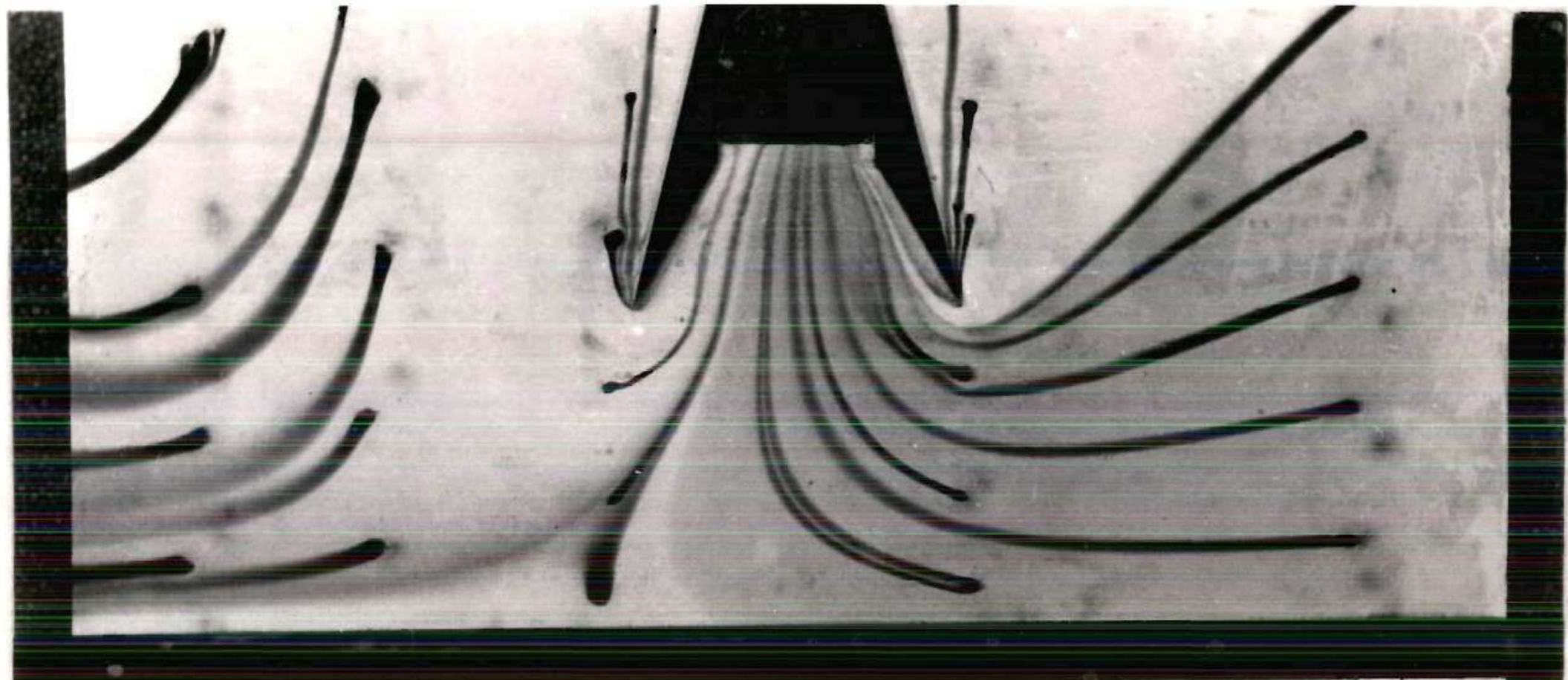}}
\subfigure[]{\includegraphics[width=0.3\textwidth]{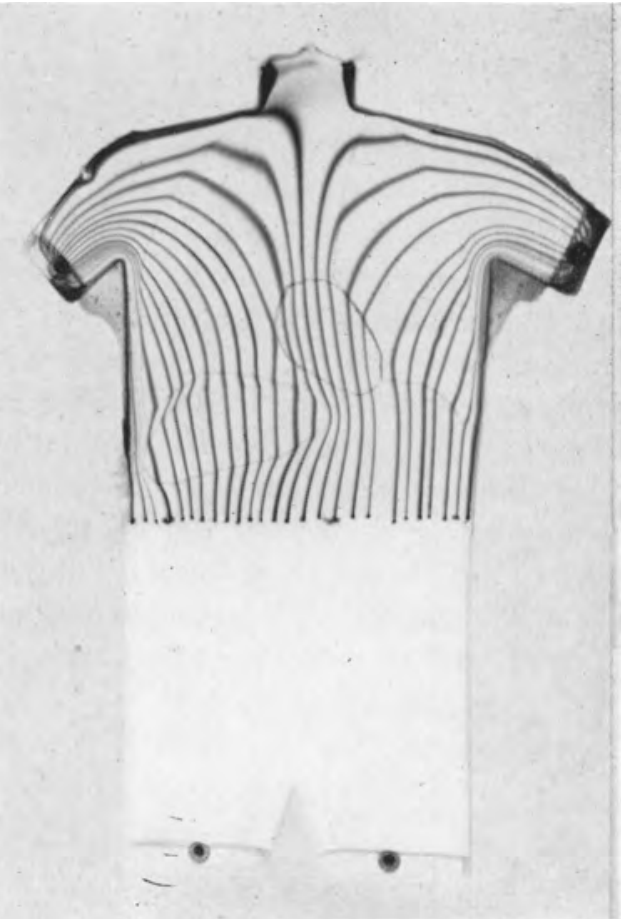}}
\caption{Applications of fluid mappers. 
(a)~Fluid mapper used in optimisation of a canopy exhaust hood in 1954~\cite{clem1954use}.
(b)~Imitation of a current flow in a human body with a thin-layer fluid flow, with domains of low permeability corresponding to lungs and liver, the fluid enters the model from the left leg and leaves the model through the arms. The experiments are conducted in 1952.
From~\cite{mcfee1952graphic}}
\label{ecg}
\end{figure}

In 1952  Moore's method was applied to study current flow for various positions of electrocardiographic leads: an outline of a human body was made of a plaster and covered with a glass plate to allow only a thin layer of fluid inside, locations of a source and sinks of fluid flow corresponded to positions of electrocardiographic
electrodes, variations of resistance of organs were modelled by varying the depth of the plaster slab~\cite{mcfee1952graphic} (Fig.~\ref{ecg}a). In 1954 a fluid mapper was evaluated  in designs of fume exhaust hoods~\cite{clem1954use}: it was possible to plot hood characteristics, stream, pressure and velocity lines with the help of the experimental fluid mapper (Fig.~\ref{ecg}b).

\graphicspath{{figs/flow/}}
\begin{figure}[!tbp]
\centering
\subfigure[0 sec]{\includegraphics[width=0.32\textwidth]{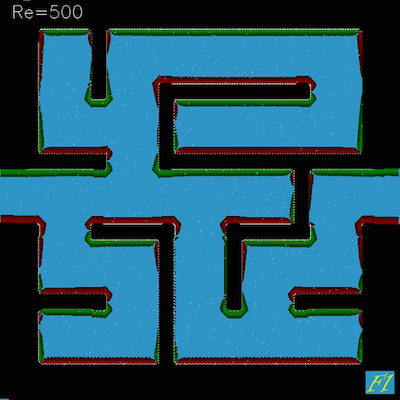}}
\subfigure[1 sec]{\includegraphics[width=0.32\textwidth]{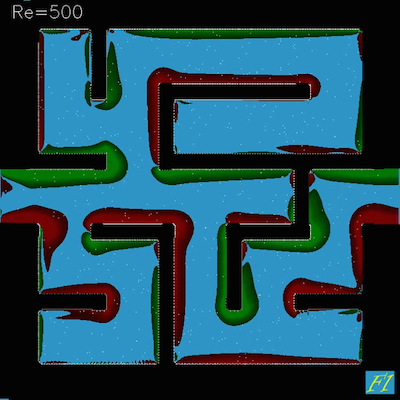}} 
\subfigure[2 sec]{\includegraphics[width=0.32\textwidth]{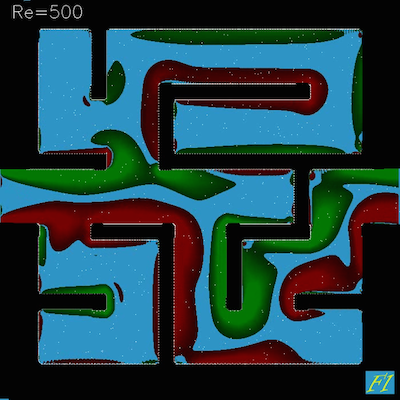}} 
\subfigure[4 sec]{\includegraphics[width=0.32\textwidth]{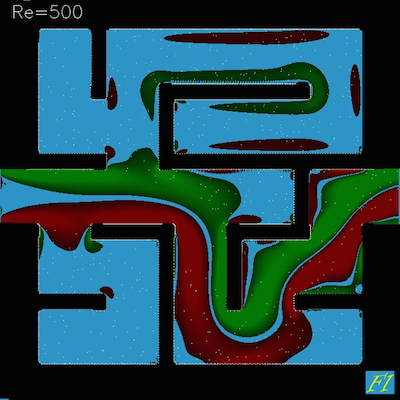}} 
\subfigure[6 sec]{\includegraphics[width=0.32\textwidth]{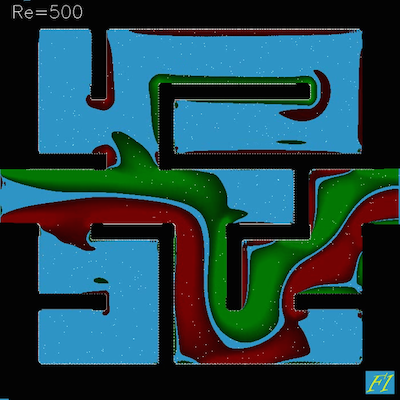}} 
\subfigure[10 sec]{\includegraphics[width=0.32\textwidth]{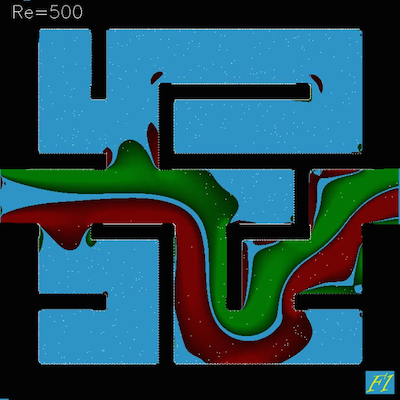}}
\graphicspath{{figs/flow/Maze_V5_Frames/}}
\subfigure[0 sec]{\includegraphics[width=0.32\textwidth]{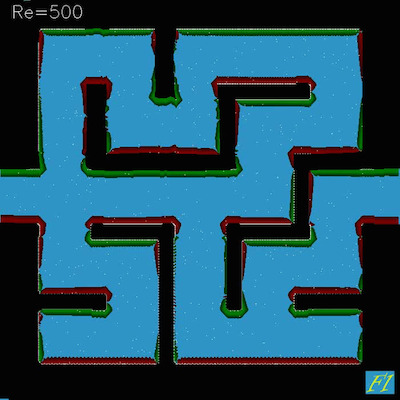}} 
\subfigure[1 sec]{\includegraphics[width=0.32\textwidth]{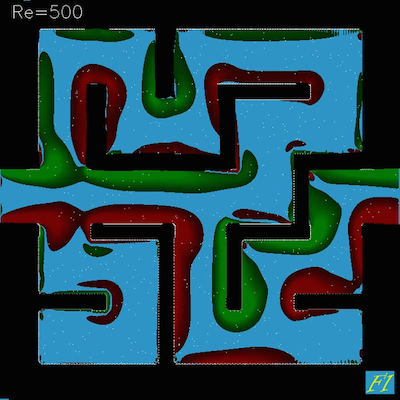}} 
\subfigure[2 sec]{\includegraphics[width=0.32\textwidth]{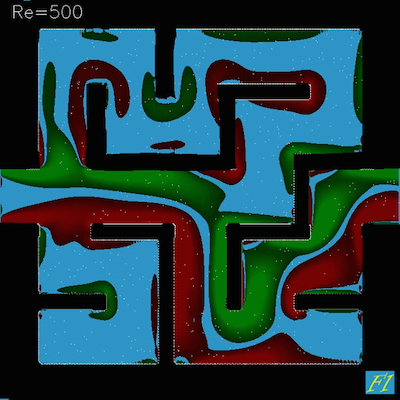}} 
\subfigure[4 sec]{\includegraphics[width=0.32\textwidth]{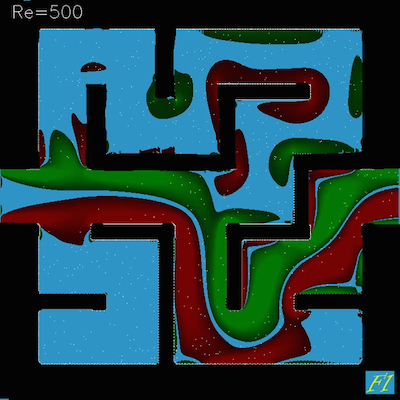}} 
\subfigure[6 sec]{\includegraphics[width=0.32\textwidth]{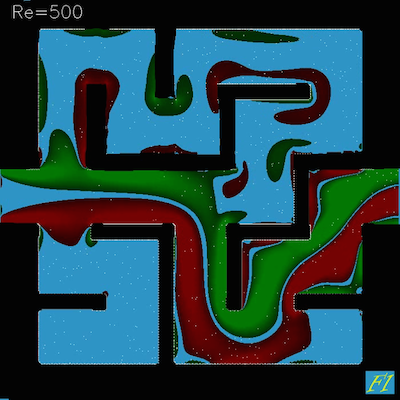}} 
\subfigure[10 sec]{\includegraphics[width=0.32\textwidth]{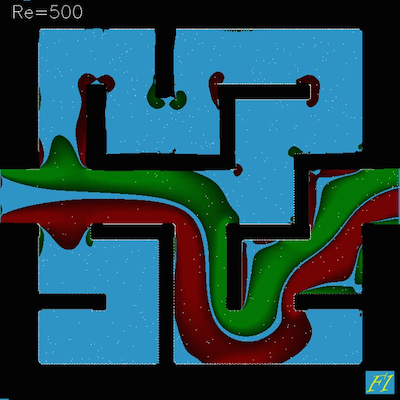}}
\caption{Fluid flow through the (a--b) labyrinth and (f--j=k) maze. Entrance is on the left, exit is on the right.
Labyrinth is generated in Maze Generator (http://www.mazegenerator.net/) and modified to maze.
Flow simulation is done in Flow Illustrator (http://www.flowillustrator.com/) for visual flow control $dt=0.01$ and Reynolds number 500. Maze is black, red coloured areas are part of fluid making clockwise rotation and green coloured areas --- conter-clockwise. See videos of these computational experiments  (https://youtu.be/FUBYr3cOoC8) (labyrinth) 
and (https://youtu.be/0jFPXBhQBS0) (maze). Time shown as per video generated by the Flow Illustrator. }
\label{reynolds500labyrinth}
\end{figure}
\graphicspath{{figs/}}

\graphicspath{{figs/milkmaze/}}
\begin{figure}[!tbp]
\centering
\subfigure[20 sec]{\includegraphics[width=0.42\textwidth]{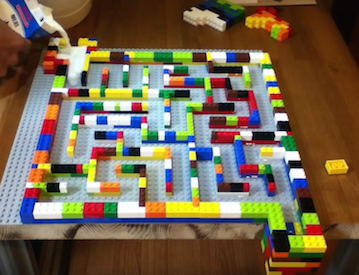}}
\subfigure[22 sec]{\includegraphics[width=0.42\textwidth]{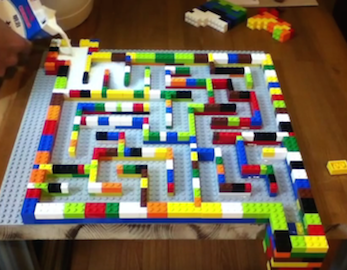}}
\subfigure[25 sec]{\includegraphics[width=0.42\textwidth]{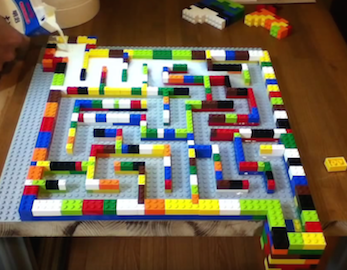}}
\subfigure[30 sec]{\includegraphics[width=0.42\textwidth]{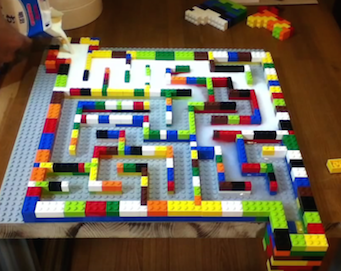}}
\subfigure[35 sec]{\includegraphics[width=0.42\textwidth]{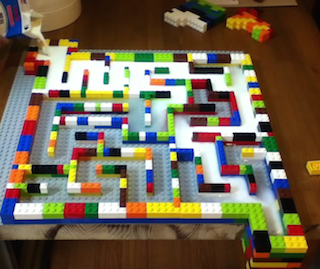}}
\subfigure[40 sec]{\includegraphics[width=0.42\textwidth]{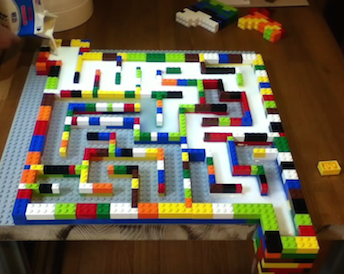}}
\caption{Snapshots of milk and water solving maze. 
Snapshots from the video of experiments by Masakazu Matsumoto \cite{Matsumoto}
who used one Lego 6177, 
one Lego 628, half-a-litre of milk and two litres of water. 
Reproduced with kind permission from Masakazu Matsumoto.}
\label{MatsumotoMaze}
\end{figure}
\graphicspath{{figs/}}

\begin{figure}[!tbp]
\centering
\subfigure[]{\includegraphics[width=0.4\textwidth]{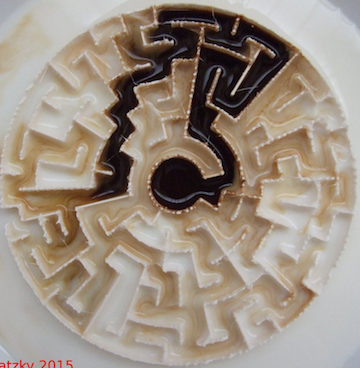}}
\subfigure[]{\includegraphics[width=0.4\textwidth]{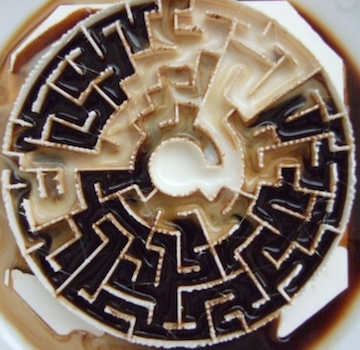}}
\caption{Labyrinth solving with coffee and milk: (a) the path is traced by coffee, (b) the path is traced by milk. }
\label{coffeemilk}
\end{figure}

First published evidence of experimental laboratory fluid maze solver is dated back to 2003. In a fluidic maze solver developed in \cite{fuerstman2003solving} a maze is the network of micro-channels. The network is sealed. Only the source site (inlet) and the destination site (outlet) are open. The maze  is filled with a high-viscosity fluid. A low-viscosity coloured fluid  is pumped under pressure into the maze, via the inlet. Due to a pressure drop between the inlet and the outlet liquids start leaving the maze via the outlet. A velocity of fluid in a channel is inversely proportional to the length of the channel. High-viscosity fluid in the channels leading to dead ends prevents the coloured low-viscosity fluid from entering the channels.  There is no pressure drop between the inlet and any of the dead ends. Portions of  the `filler' liquid leave the maze. They are gradually displaced by the colour liquid. The colour liquid travels along maximum gradient of the pressure drop, which is along a shortest path from the inlet to the outlet. When the coloured liquid fills  the path the viscosity along the path decreases. This leads to an increase of the liquid velocity along the path.  The shortest path --- least hydrodynamic resistance \index{hydrodynamic resistance} path ---  from the inlet to the outlet is represented by channels filled with coloured fluid. Visualisation of the fluid flow indicating a shortest path in a maze is shown in Fig.~\ref{reynolds500labyrinth}. Fluids solve mazes at any scale, not just micro-fluidics, as has been demonstrated by Masakazu Matsumoto, where water explores the maze and milk traces the shortest path (Fig.~\ref{MatsumotoMaze}) and in our own experiments with milk and coffee in a labyrinth (Fig.~\ref{coffeemilk}).

\section{Electrical mappers. Shortest path is a path of the least electrical resistance.}

\graphicspath{{figs/electrical/}}
\begin{figure}[!tbp]
\centering
\subfigure[]{\includegraphics[width=0.5\textwidth]{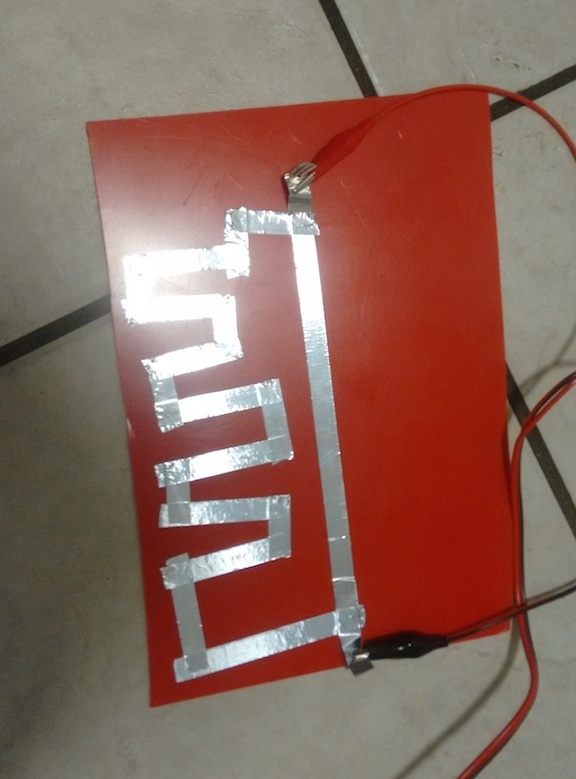}}
\subfigure[]{\includegraphics[width=0.46\textwidth]{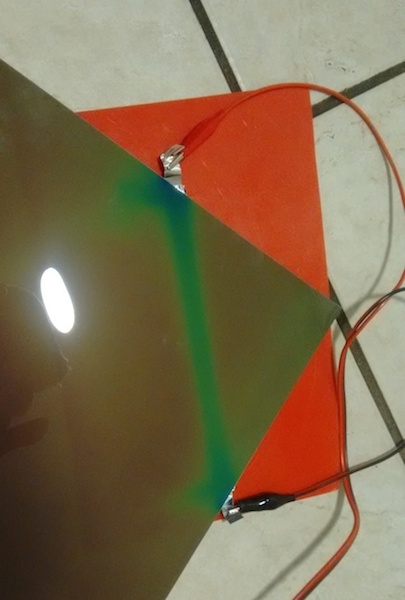}}
\caption{Calculating a shortest path with electrical current.
(a)~Circuit.
(b)~Visualisation of a shortest path with the temperature sensitive liquid crystal sheet (Edmund Optics Inc., USA). Current applied through is 3.2A.  
See Chapter by Simon Ayrinhac \cite{adamatzkySPbook} for more details on electrical-current based path finding.
}
\label{electrical}
\end{figure}
\graphicspath{{figs/}}

Approximation of a collision-free path \index{collision-free path} with a network of resistors \index{resistor} was first proposed in \cite{tarassenko1991analogue, tarassenko1991parallel}. 
A space is represented as a resistor network, obstacles are insulators. An electrical power source is connected to the destination and the source sites.  The destination site is the electrical current source. Current flows in the network but does not enter obstacles. A path can be traced by a gradient descent in electrical potential.  That is for each node a next move is selected by measuring the voltage difference between the current node and each of its neighbours, and moving to the neighbours which shows maximum voltage.  As shown by Simon Ayrinhac (originally in~\cite{ayrinhac2014electric}), a shortest path can be visualised without discretisation of the space. A maze is filled with a continuous conductive material. Corridors are conductors, walls are insulators. An electrical potential difference is applied between the source and the destination sites. The electrical current `explores' all possible pathways in the maze.  An electrical current is stronger along the shortest path.  Local temperature in a locus of a conducting material is proportional to a current strength through  this locus.  A temperature profile can be visualised with thermal camera~\cite{ayrinhac2014electric} or glow-discharge~\cite{reyes2002glow} or
temperature sensitive liquid crystal sheets (Fig.~\ref{electrical}).

\section{Diffusion mappers. Shortest path is a path of the steepest gradient of chemoattractants}

A source of a diffusing substance is placed at the destination site.  After the substance propagates all over the maze a concentration of the substance develops.  The concentration gradient is steepest towards the source of the diffusion. Thus starting at any site of the maze and following the steepest gradient one can reach the source of the diffusion. The diffusing substance represents one-destination-many-sources shortest paths. To trace a shortest path from any site, we place a chemotactic agent at the site and record its movement towards the destination site. There are three experimental laboratory prototypes of visualising a shortest path in a diffusion field: by using travelling droplets, crawling epithelial cells and growing slime mould. 

A path along the steepest gradient of potassium hydroxide  has been visualised  by Istv{\'a}n Lagzi and colleagues with a droplet of a mineral oil or dichloromethane mixed with 2-hexyldecanoic acid~\cite{lagzi2010maze} (see Chapter by Jitka \v{C}ejkov\'{a} et al in \cite{adamatzkySPbook}). Daniel Irimia and colleagues used epithelial cells \index{epithelial cell}  to visualise the steepest gradient of the epidermal growth factor~\cite{scherber2012epithelial} (see Chapter by Daniel Irimia in \cite{adamatzkySPbook}). Let us discuss our own experiments on visualising a path in a maze with slime mould.

\begin{figure}[!tbp]
\centering
\subfigure[]{\includegraphics[width=0.45\textwidth]{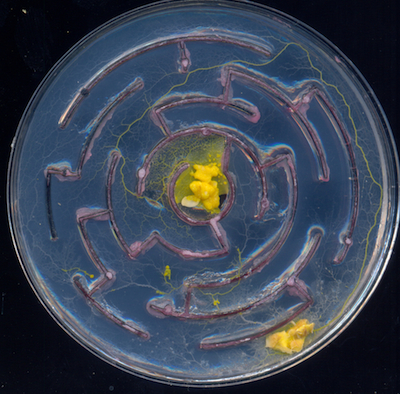}}
\subfigure[]{\includegraphics[width=0.45\textwidth]{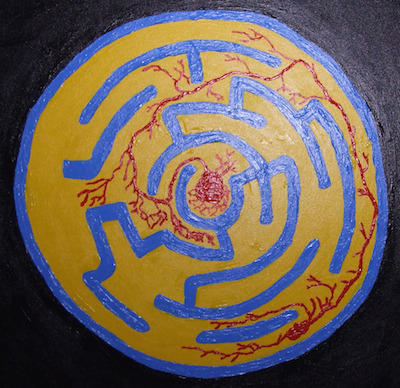}}
\caption{A path between central chamber of exit of a maze is represented by thickest protoplasmic tube. (a)~Photo of experimental setup with the maze solved. (b)~Painting of the setup where the path to the maze's central chamber is more visible.}
\label{phymaze}
\end{figure}

\graphicspath{{figs/diffusion/}}
\begin{figure}[!tbp]
\centering
\subfigure[$t=101$]{\fbox{\includegraphics[width=0.24\textwidth]{Diffusion000101}}}
\subfigure[$t=202$]{\fbox{\includegraphics[width=0.24\textwidth]{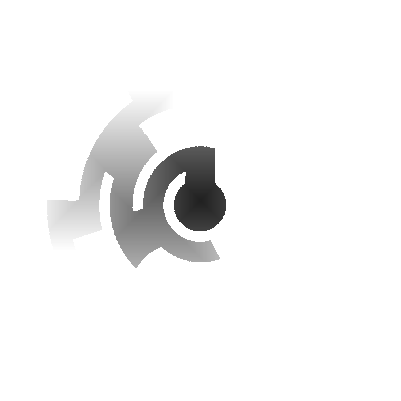}}}
\subfigure[$t=303$]{\fbox{\includegraphics[width=0.24\textwidth]{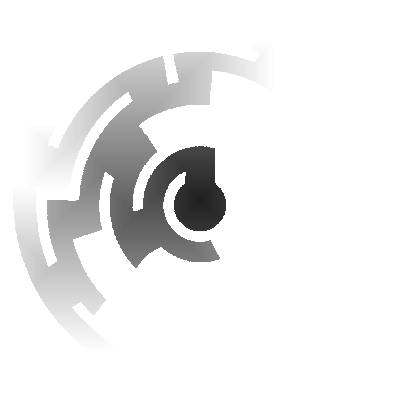}}}
\subfigure[$t=404$]{\fbox{\includegraphics[width=0.24\textwidth]{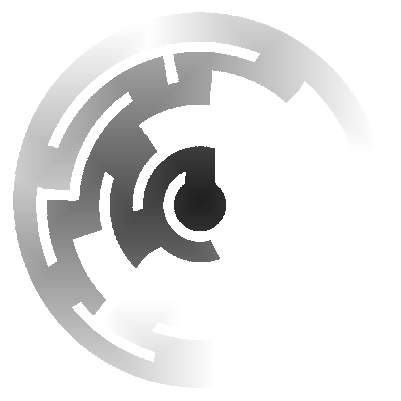}}}
\subfigure[$t=505$]{\fbox{\includegraphics[width=0.24\textwidth]{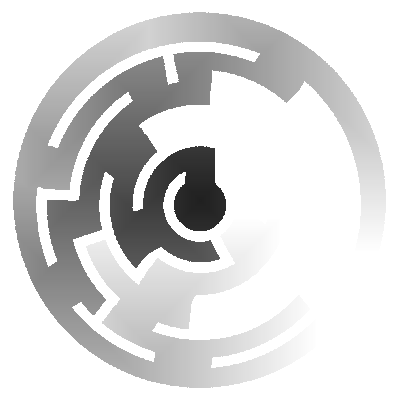}}}
\subfigure[$t=606$]{\fbox{\includegraphics[width=0.24\textwidth]{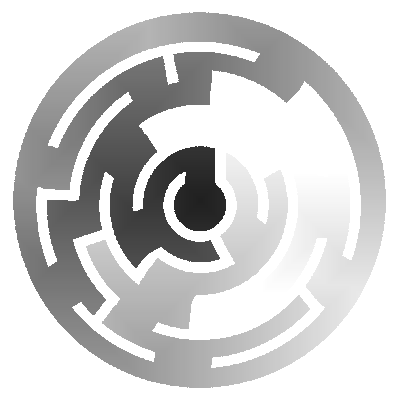}}}
\caption{Numerical simulation of a diffusing chemo-attractant.  
The grey-level is proportional to a concentration of the chemo-attractant. 
Time steps indicated are iteration of numerical integration.
See details in~\cite{adamatzky2012slime}.
}
\label{diffusion}
\end{figure}

\graphicspath{{figs/excitation/}}
\begin{figure}[!tbp]
\centering
\subfigure[$t=807$]{\fbox{\includegraphics[width=0.24\textwidth]{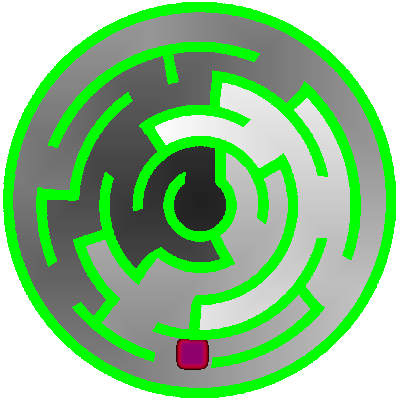}}}
\subfigure[$t=1307$]{\fbox{\includegraphics[width=0.24\textwidth]{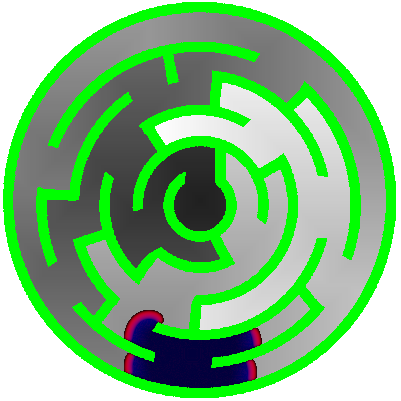}}}
\subfigure[$t=1807$]{\fbox{\includegraphics[width=0.24\textwidth]{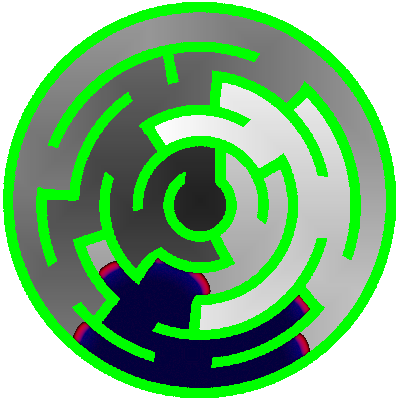}}}
\subfigure[$t=2307$]{\fbox{\includegraphics[width=0.24\textwidth]{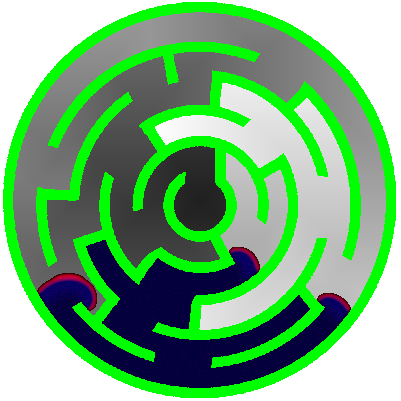}}}
\subfigure[$t=2807$]{\fbox{\includegraphics[width=0.24\textwidth]{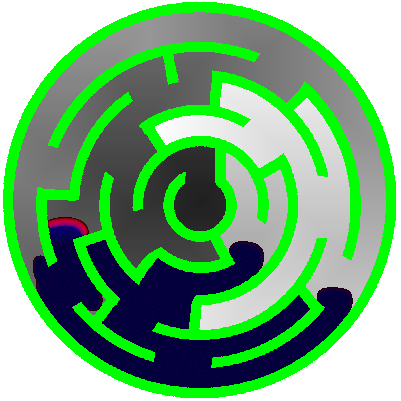}}}
\subfigure[$t=4807$]{\fbox{\includegraphics[width=0.24\textwidth]{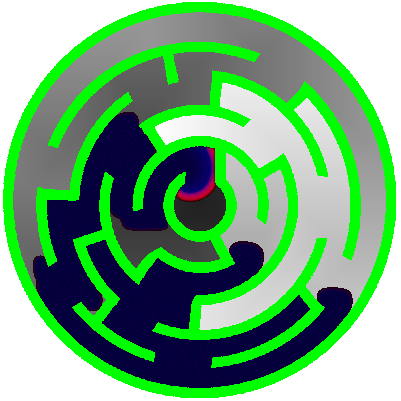}}}
\caption{ Numerical simulation of the gradient tracing by the slime mould.  Active growing zone is shown by red colour. 
The slime mould's body is shown by blue colour. See details of the model in ~\cite{adamatzky2012slime}.}
\label{diffusion}
\end{figure}

The slime mould \index{slime mould} maze solver based on chemo-attraction is proposed in~\cite{adamatzky2012slime}. An oat flake is placed in the destination site. The slime mould \emph{Physarum polycephalum} \index{Physarum polycephalum} is inoculated in the source site. The oat flakes, or rather bacterias colonising the flake, release a chemoattractant. The chemo-attractant diffuses along the channels (Fig.~\ref{diffusion}).  The slime mould explores its vicinity by branching protoplasmic tubes into  openings of nearby channels.  When a wave-front of diffusing attractants reaches the slime mould, the cell halts its lateral exploration.  The slime mould develops an active growing zone propagating along the gradient of the  attractant's diffusion. The problem is solved when the slime mould reaches the source site. The thickest tube represents the shortest path between the destination site and the source site (Fig.~\ref{phymaze}). Mechanisms of tracing the gradient by the slime mould are confirmed via numerical simulation a two-variable Oregonator partial-differential equations in a two-dimensional space (Fig.~\ref{diffusion}).    Not only nutrients can be placed at the destination site but  any volatile substances that attract the slime mould, e.g. roots of the medicinal plant \emph{Valeriana officinalis} \cite{ricigliano2015plant}. Note that in our experiments reported in~\cite{adamatzky2012slime} slime mould did not calculate the shortest path inside the maze but just one of the paths, while Oregonator based model always produces the shortest path.

\section{Thoughts on inconclusive experiments}

\graphicspath{{figs/}}
\begin{figure}[!tbp]
\centering
\subfigure[]{\includegraphics[width=0.3\textwidth]{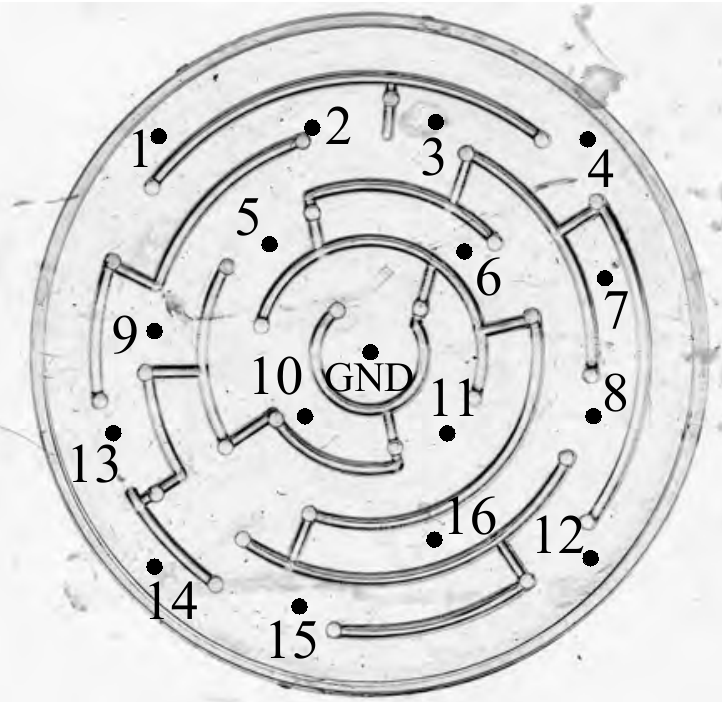}}
\subfigure[]{\includegraphics[width=0.3\textwidth]{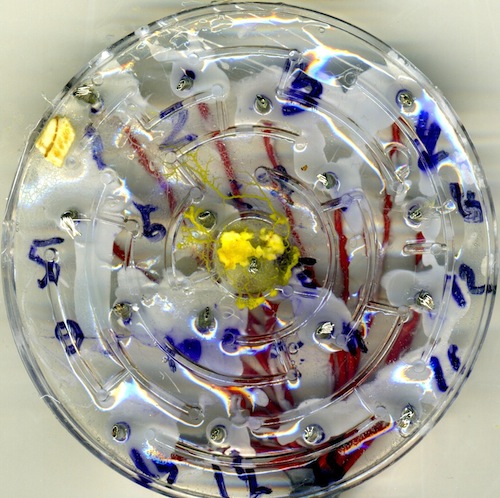}}
\subfigure[]{\includegraphics[width=0.3\textwidth]{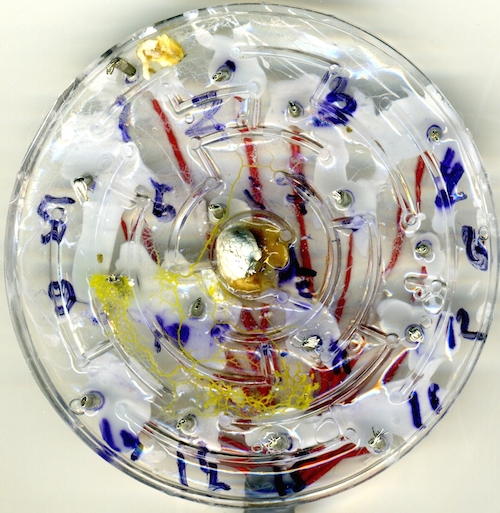}}
\subfigure[]{\includegraphics[width=0.9\textwidth]{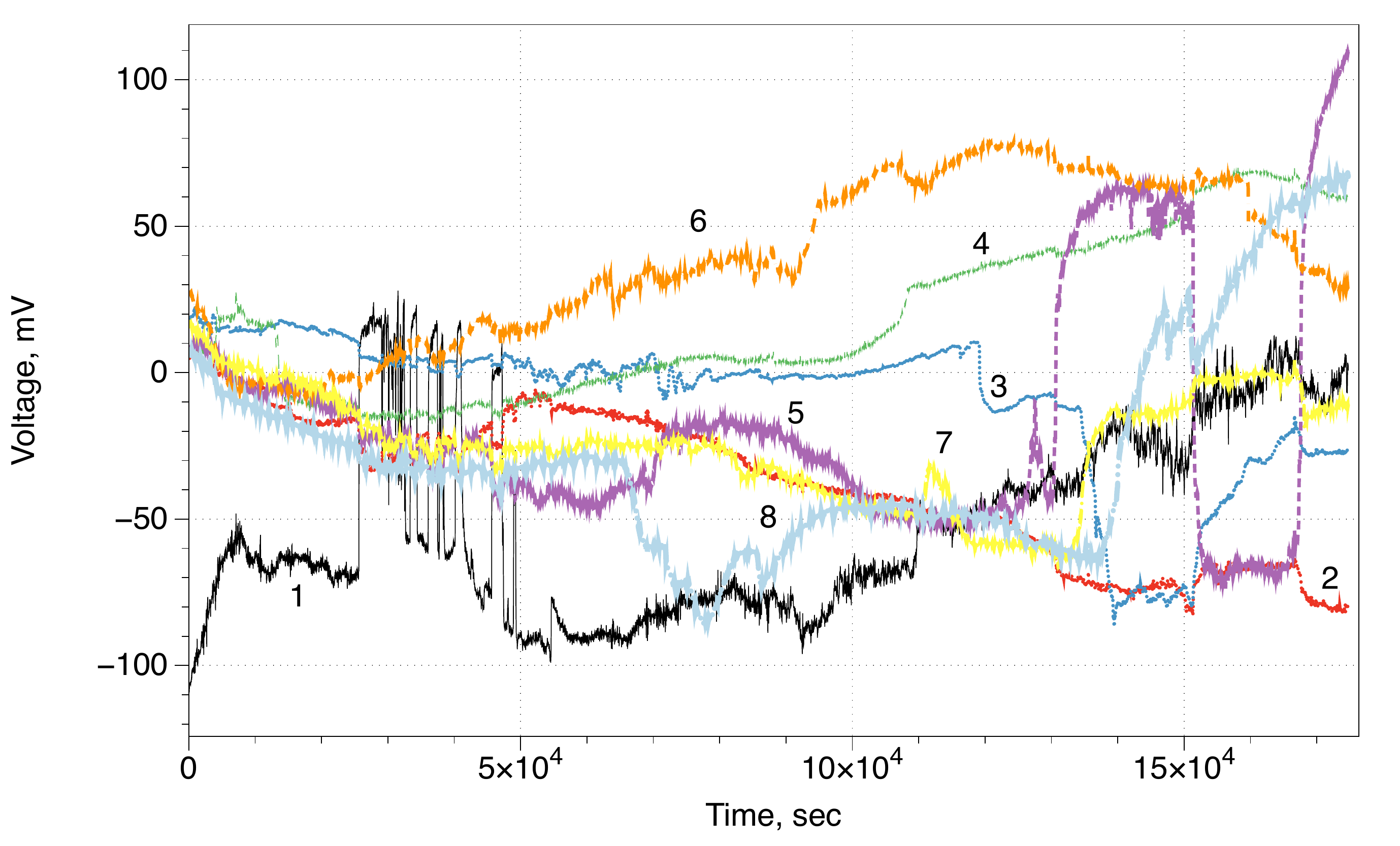}}
\subfigure[]{\includegraphics[width=0.9\textwidth]{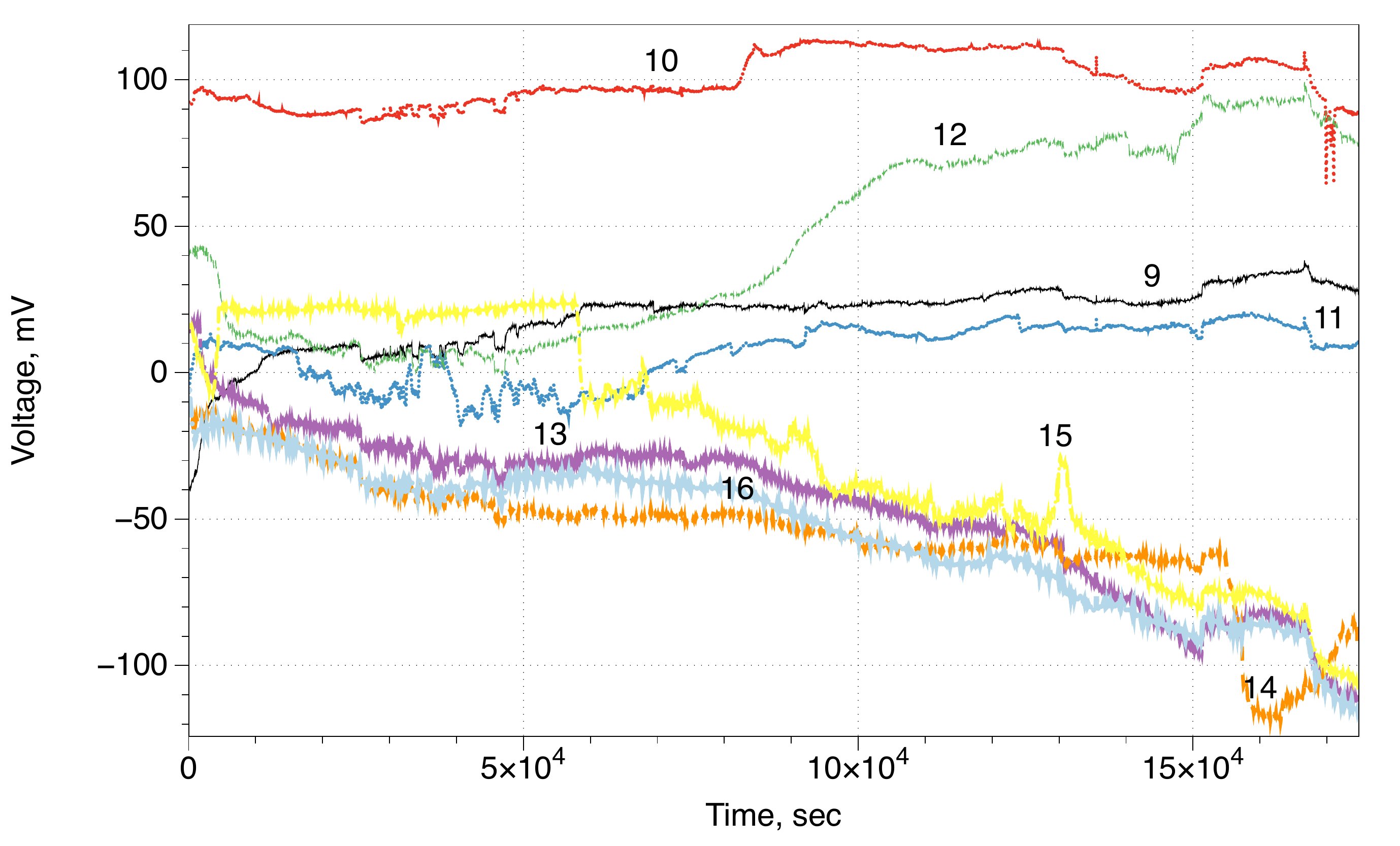}}
\caption{Illustrations of experiments on electrical activity of slime mould \emph{Physarum polycephalum} during maze solving: 
(a) positions of electrodes, reference electrode is labeled GND, 
(b) two days after inoculation, 
(c) three days after inoculation,
(d) electrical activity recorded on channels 1 to 8, sampling rate is one per second,
(e) electrical activity recorded on channels 9 to 16, plots of electrical potential on electrodes are shown by different colours and also numbered.
}
\label{physarummazelectrodes}
\end{figure}

In 2009 we attempted to understand what is going on in the slime mould's `mind' when it traces gradients of chemoattractants. We positioned 16 electrodes at the bottom of a plastic maze (Fig~\ref{physarummazelectrodes}a) with reference electrode in the central chamber, poured some agar above,  inoculated the slime mould in the central chamber and placed an oat flake at the outer channel of the maze. Configurations of the slime mould growing in the maze are shown  Fig~\ref{physarummazelectrodes}bc.  Electrical potential differences between each of 16 electrodes and the reference electrode recorded during several days are shown in Fig.~\ref{physarummazelectrodes}de. We found that active growing zones of the slime mould show a higher level of the electrical potential difference and there are signs of  an apparent communication between the zones performing parallel search at different parts of the maze. However, it still remains unclear how exactly `suppression' of growing zones propagating along the longest routes is implemented.

\begin{figure}[!tbp]
\centering
\subfigure[]{\includegraphics[width=0.42\textwidth]{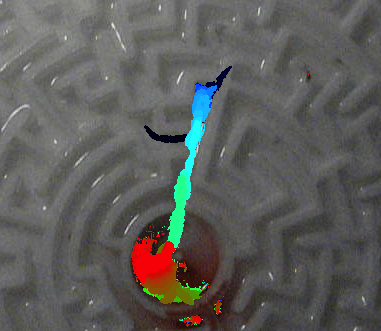}}
\subfigure[]{\includegraphics[width=0.487\textwidth]{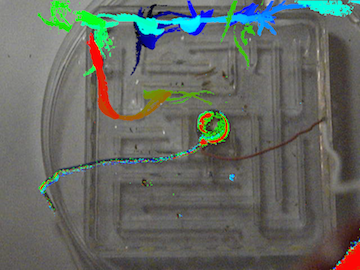}}
\caption{A leech partially solves maze. 
(a)~Trajectory of the leech approaching the target, blue pixels show starting position of a leech, red pixels --- final position.
(b)~Trajectory of a leech in a template with vibrating motor.}
\label{leaches}
\end{figure}

A spectrum of leeches' behaviour traits is extensively classified~\cite{dickinson1984feeding}.   A leech positions itself at the water surface in resting state. The leech swims towards the source of a mechanical or optical stimulation. The leech stops swimming when it comes into contact with any geometrical surface. Then, the leech explores the surface by crawling. When a leech finds a warm  region the leech bites.  We attempted to solve a maze, a template printed of nylon and filled with water, with young leeches \emph{Hirudo verbana} (see details of experimental setup in \cite{adamatzky2015exploration}). We tried fresh blood, temperature and vibration as  sources of physical stimuli which would attract leeches to the target site. The leeches did not show attraction to the blood when placed over 5~cm away from the source. Being placed in the proximity of the source the leeches crawled or swam to the target site (Fig.~\ref{leaches}a). We have also conducted scoping experiments with leeches in presence of thermal gradients. To form the gradient we immersed, by 5~mm, a tip of a soldering iron, heated to 40$^o$C, in the water inside a central chamber of the maze. In half of the experiments, leeches escaped from the template, in a quarter of the experiments leeches moved to the domains proximal to the source of a higher temperature and in a quarter of experiments leeches moved towards the source of thermal stimulation. Trajectories of the leeches movement in the presence of a source of vibration did not show any statistically significant preference towards movement into areas with highest level of vibration, in some cases a leech was moving towards the vibrating motor from the start of an experiment but then swam or crawled  away (Fig.~\ref{leaches}b). Inconclusive results with vibration-assisted maze solving could be due to a reflection of waves at maze walls.

\begin{figure}[!tbp]
\centering
\subfigure[]{\includegraphics[width=0.7\textwidth]{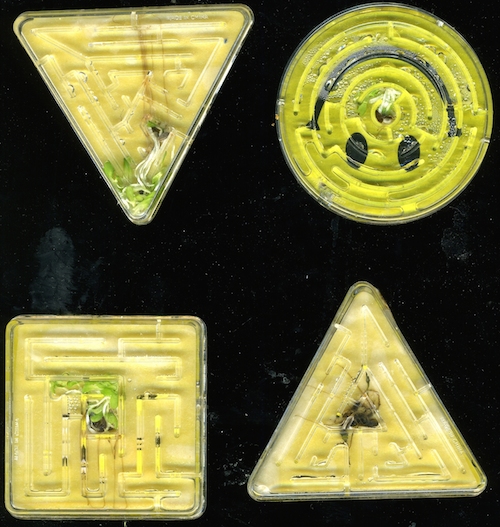}}
\subfigure[]{\includegraphics[width=0.45\textwidth]{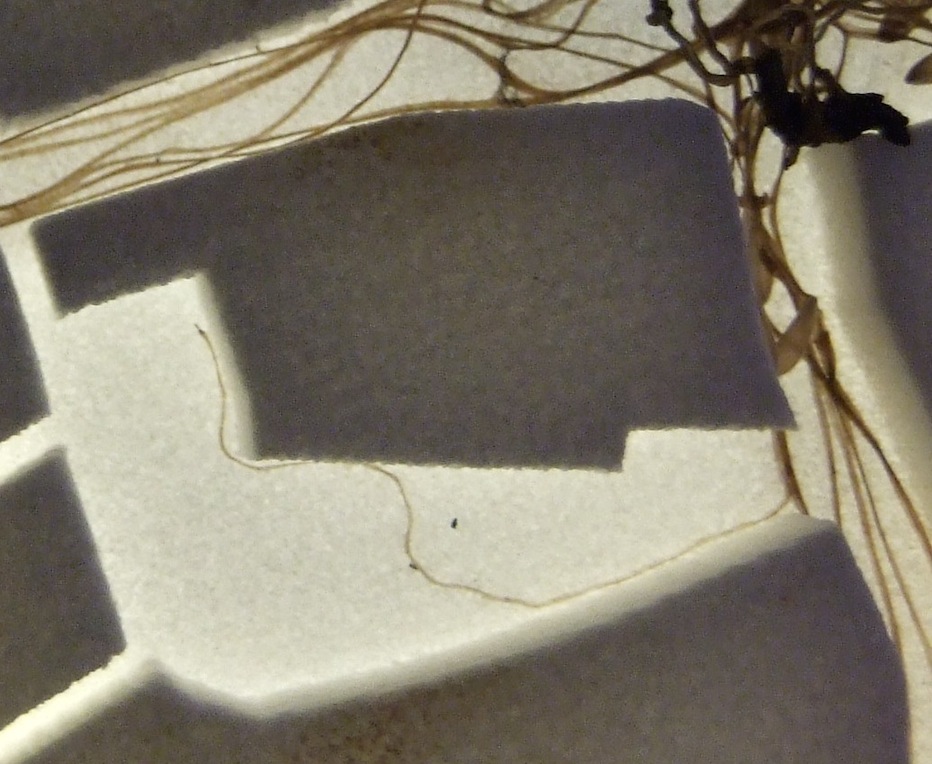}}
\subfigure[]{\includegraphics[width=0.45\textwidth]{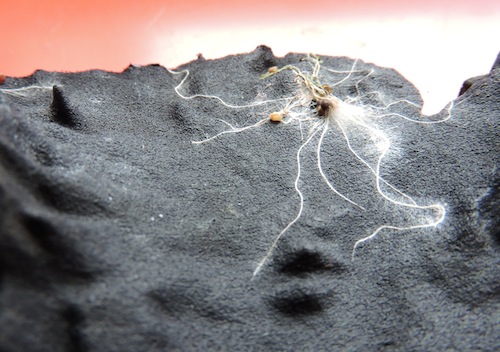}}
\caption{(a) Lettuce seedlings grow inside plastic mazes. 
(bc)~Roots propagate in a 3D model of Bristol. 
(d)~Maize roots navigate around elevations on 3D template of USA.
}
\label{plants}
\end{figure}

We have undertaken a few scoping experiments on plants navigating mazes guided only by gravity force and physical structure of a maze, see  illustrations in Fig.~\ref{plants}a. When seeds are placed in or near a central chamber of a maze their roots somewhat grow towards the exit of the labyrinth.  However, they often become stuck midway and rarely reach the exit.  Yokawa and colleagues~\cite{yokawa2014binary} demonstrated that by using volatiles it is possible to navigate the roots in simple binary mazes, more complicated mazes have not been tested.  Few more experiments on a collision-free path approximation by plant roots have been done on a 3D templates of Bristol (UK) city and USA.  The seeds of  lettuce, in experiments with a template of Bristol, were placed in large open spaces, corresponding to squares. The templates were kept in a horizontal position.  We found that root apexes prefer wider streets, they rarely enter side streets (Fig.~\ref{plants}b).  Potential prototypes of shortest path solvers with roots could be the case of future studies, at this moment we only know that roots navigate around obstacles (Fig.~\ref{plants}b) and elevations (Fig.~\ref{plants}c).  

\section{Conclusion}

To solve a maze we need a mapper and a tracer. The tracer's role is straightforward, we would say easy, just follow a map made by the mapper. This is the mapper who does all `computation'. Does it? And here we come to a disturbing thought that a computation exists only in our mind. Nature does not compute. It is us who invented the concept of computation.  As Stanley Kubrick told in his interview to ``Playboy'' magazine in 1968~\cite{kubrick2001stanley}: 

\begin{quote}
The most terrifying fact about the universe is not that it is hostile but that it is indifferent; but if we can come to terms with this indifference and accept the challenges of life within the boundaries of death --- however mutable man may be able to make them --- our existence as a species can have genuine meaning and fulfilment.
\end{quote}

Designing and re-designing experimental laboratory prototypes of unconventional computing devices might be our way to cope with the Nature's indifference.

\bibliographystyle{plain}
\bibliography{biblioadamatzkysp}

\end{document}